\documentclass[%
 reprint,
 amsmath,amssymb,
 prb,
 aps,
]{revtex4-1}

\usepackage{graphicx}
\usepackage{dcolumn}
\usepackage{bm}
\usepackage{overpic}
\usepackage{xcolor}

\begin{document}
\preprint{APS/123-QED}

\title{Double quantum-dot engine fueled by entanglement between electron spins}

\author{Martin Josefsson}
\author{Martin Leijnse}
\affiliation{Solid State Physics and NanoLund, Lund University, Box 118, S-221 00 Lund, Sweden}

\date{\today}
\begin{abstract}
The laws of thermodynamics allow work extraction from a single heat bath provided that the entropy decrease of the bath is compensated for by another part of the system. We propose a thermodynamic quantum engine that exploits this principle and consists of two electrons on a double quantum dot (QD). The engine is fueled by providing it with singlet spin states, where the electron spins on different QDs are maximally entangled, and its operation involves only changing the tunnel coupling between the QDs. Work can be extracted since the entropy of an entangled singlet is lower than that of a thermal (mixed) state, although they look identical when measuring on a single QD. We show that the engine is an optimal thermodynamic engine in the long time limit. In addition, we include a microscopic description of the bath and analyze the engine's finite time performance using experimentally relevant parameters. \\
\end{abstract}


\maketitle
\section{\label{sec:introduction} Introduction}
The rapidly developing field of quantum thermodynamics~\cite{vinjanampathy2016quantum, binder2019thermodynamics} attempts to combine the laws of quantum mechanics with those of classical thermodynamics. One of the central goals is to investigate the performance of thermodynamic engines operating in the quantum regime. Quantum effects can, in principle, boost the performance of thermodynamic engines -- see, e.g., the theoretical works in Refs.~\onlinecite{bengtsson2018quantum, scully2003extracting, dillenschneider2009energetics, rossnagel2014nanoscale, del2014more, binder2015quantacell} -- but limitations set by the second law of thermodynamics, such as the Carnot efficiency of a heat engine, still always hold.

The second law can appear to be violated in setups where some other resource, beyond heat, is in fact consumed by the engine, the entropy content of which should be accounted for. This resource can, for example, be a non-thermal state in the reservoirs,~\cite{scully2003extracting, dillenschneider2009energetics, rossnagel2014nanoscale, del2014more, abah2014efficiency, binder2015quantacell, niedenzu2016operation, whitney2019non} or information about the working "gas" in Maxwell's demon-type engines.~\cite{maruyama2009colloquium} In cyclic versions of demon-type engines the acquired information needs to be erased, a process that requires expenditure of resources. The famous result of Landauer\cite{landauer1961irreversibility} states that the minimum energy required to erase the information of a bit (e.g. put it in the $0$ state) is on average $k_B\tau\ln 2$,  where $k_B$ is the Boltzmann constant and $\tau$ the temperature of the environment, although the resource spent need not be in the form of energy.~\cite{vaccaro2011information, barnett2013beyond} Several experiments have confirmed this bound.~\cite{berut2012experimental, jun2014high, hong2016experimental} Conversely, the maximum amount of work that can on average be extracted from a single bath using the information content of a bit is $k_B\tau\ln 2$, which has been utilized to make on-chip compatible, information-driven, generators and refrigerators.~\cite{koski2014experimental, koski2015chip, chida2017power, cottet2017observing, naghiloo2018information, masuyama2018information} If the system instead has $d$ states, the maximum energy converted is $k_B\tau\ln d$. In the quantum regime the same bounds remain valid,~\cite{alicki2004thermodynamics, del2011thermodynamic, frenzel2014reexamination} and Landauer's principle has recently been experimentally tested.~\cite{ gaudenzi2018quantum, yan2018single}
\begin{figure}[h!]
	\centering			
	\includegraphics[width=0.48\textwidth, trim={0 340 0 0},clip]{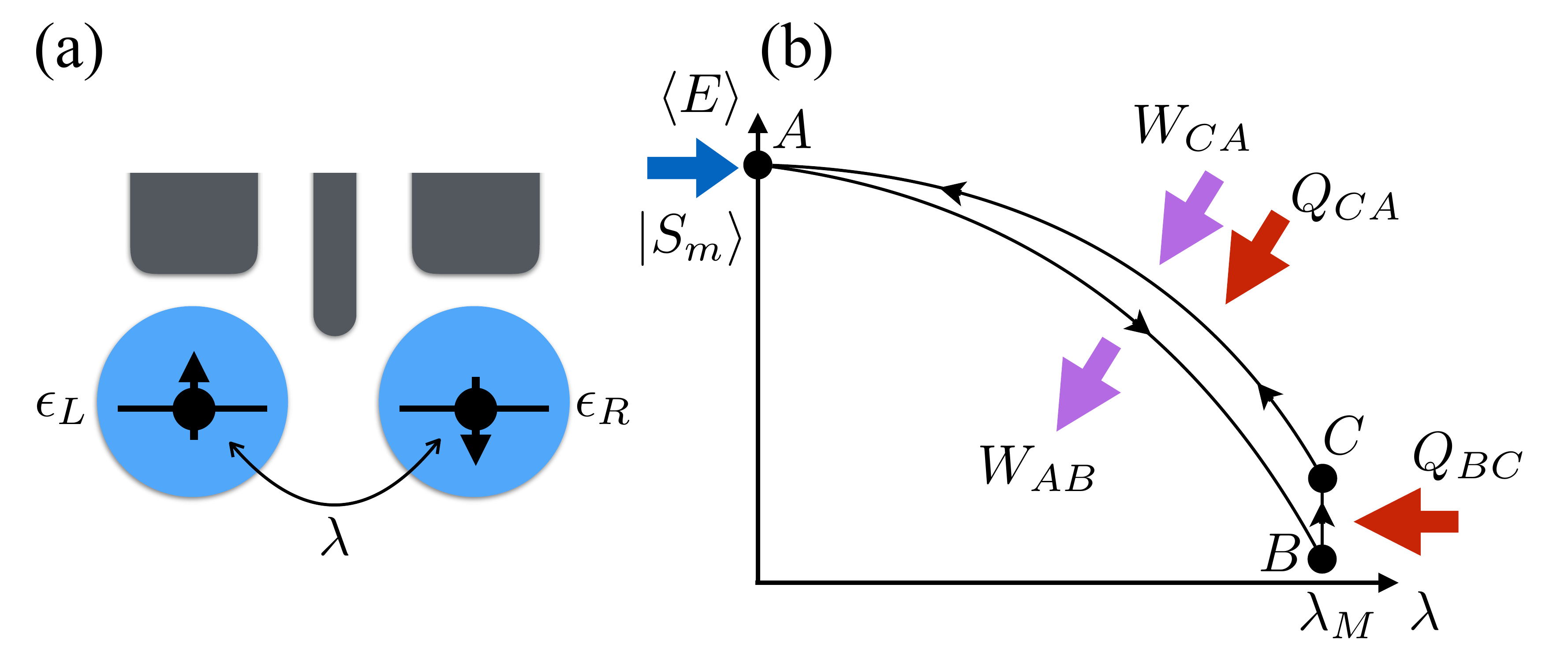}
	\caption{(a) Sketch of the double QD setup. QD $L,R$ has orbital energy $\epsilon_{L,R}$ and the QDs are coupled by tunneling with amplitude $\lambda$. The gray structures on top represent plunger gates used to control $\epsilon_{L,R}$ (which remain fixed during operation of our engine) and a pincher gate which controls $\lambda$ (varied in the engine cycle). (b) Sketch of working cycle for the engine, showing the expectation value of the double QD energy $\langle E \rangle$ as a function of $\lambda$. $W$/$Q$ represent work performed/heat absorbed (see details in Sec.~\ref{sec:basic}) and the blue arrow at point A emphasizes that we need to supply a singlet state at the start of each cycle.}		
	\label{fig:setup}
\end{figure}

In this paper we propose a quantum mechanical version of a single bath heat engine and investigate its time-dependence, revealing how to optimize work and power output under realistic experimental conditions. The engine is based on the singlet and triplet states of two coupled spins in a double quantum dot (QD), see Fig.~\ref{fig:setup}. Its cycle requires changing only a single parameter between two values, namely the inter-QD tunnel coupling which controls the energy difference between the lowest singlet state and the triplets. The proposed engine is quantum in the sense that its cycle starts with the system in a maximally entangled two-electron state, and after work has been extracted the entropy of the electrons is increased by "consuming the entanglement" (converting the initial coherent state into a thermal mixed state). Therefore, entanglement plays the role of the engine's fuel. The initial and final states will have exactly the same energy and, in fact, any measurement of the spin on a single QD will give identical random results for the two states. Nonetheless, the initial entangled state has a lower entropy than the thermal mixed state, allowing us to use it as a resource. We note, however, that our proposal is not a generic recipe for using entanglement to extract work, and that similar protocols for work extraction are possible in other systems where entanglement is not needed, but in this system it is crucial.

Both the physical setup and the needed operations are closely related to the well established platform of spin qubits in semiconductor QDs.~\cite{loss1998quantum, schliemann2001double, petta2005coherent, hanson2007spins} Our proposed engine provides a path towards near-term demonstrations of fundamental principles of quantum thermodynamics. In addition, it could be of great practical importance for making use of quantum resources produced in future quantum computers or simulators by consuming left-over entangled states to extract work and cool the quantum circuits.
\section{\label{sec:model} Model}
We consider a double QD as sketched in Fig.~\ref{fig:setup}(a), described by the Hamiltonian
\begin{eqnarray}\label{eq:HDQD}
	H_{DQD} &=& \sum_{i, \sigma} \epsilon_i n_{i \sigma} + \sum_i U_i n_{i \uparrow} n_{i \downarrow} + U_{LR} n_L n_R \nonumber \\
		&+& \lambda \sum_{i \neq j, \sigma} c_{i \sigma}^\dagger c_{j \sigma} + h.c.,  
\end{eqnarray}
where $c_{i\sigma}^\dagger$ creates an electron on QD $i = L,R$ with spin projection $\sigma = \uparrow, \downarrow$, $n_{i \sigma} = c_{i\sigma}^\dagger c_{i\sigma}$, $n_i = \sum_{\sigma} n_{i \sigma}$, $U_i$ is the intra-QD Coulomb repulsion on QD $i$, $U_{LR}$ is the inter-QD Coulomb repulsion, and $\lambda$ is the strength of the hybridization between the QDs. Throughout the paper we set $\epsilon_L=\epsilon_R$, $U_L=U_R$ and $U_{LR}=0$, although our results would be qualitatively the same also for $U_{LR}> 0$ and $U_L\ne U_R$, as long as $U_L,U_R\gg U_{LR}$. Under these circumstances the two-electron eigenstates are three singlets, which we denote $S_m$, $S_\pm$, and three triplets, $T_0$ and $T_{\pm}$. We distinguish the singlets by their behavior at small $\lambda$, where $S_m \approx S(1,1)$ (one electron in each QD) and $S_\pm \approx \frac{1}{\sqrt{2}}\left (S(2,0)\pm S(0,2)\right )$ (two electrons on the same dot), see Appendix. Furthermore, we assume that the double QD spin states thermalize at temperature $\tau$ on some characteristic time-scale.   
\section{\label{sec:basic} Basic principle and ideal performance}
Here we first describe the basic principle of operation of our engine and calculate the work output for two different cycles, denoted $(i)$ and $(ii)$, under ideal conditions and without considering the details of the couplings to the thermal bath. In Sect.~\ref{sec:details} we will include a microscopic model of the bath and derive estimates of the output power under realistic operating conditions.

Both cycles start at point A [see Fig.~\ref{fig:setup}(b)] with the two QDs uncoupled, $\lambda=0$. Then $S_m = S(1,1)$, and $S_m$ and $T$ are degenerate while the remaining two states are split off by the large intra-QD Coulomb interaction, and will not be a part of further analysis. If a magnetic field is applied, only $S_m$ and $T_0$ will remain degenerate and our findings will be slightly modified. We assume that we start with the QDs being in the state $S_m$; this state may be left over from a quantum computational operation, it may have been intentionally prepared by a measurement in the singlet-triplet basis,~\cite{petta2005coherent, barthel2009rapid} or provided by coupling the double QDs to a superconductor.~\cite{recher2001andreev, hofstetter2009cooper} What is important is that the initial pure state should be seen as a resource that will be consumed by the engine. We note that if the singlet was obtained by a projective measurement of a thermal state our engine could be seen as a quantum mechanical version of the Szilard engine.\cite{szilard1929entropieverminderung}

In the next step an energy gap between $S_m$ and $T$ is created and the system is taken from A to B in Fig.~\ref{fig:setup}(b), which is done by increasing the hybridization to $\lambda = \lambda_M$. This lowers the energy of $S_m$ because of mixing with $S_+$, while the energy of $T$ remains unchanged, see Appendix. This step should ideally be done fast enough such that there are no transitions to the $T$ states induced by the environment (adiabatic in the classical thermodynamic sense), but slow enough that there are no diabatic transitions to the high energy singlets (adiabatic in the quantum sense). The decrease in energy corresponds to electrical work performed on the gates controlling $\lambda$, $W_{AB} = E_{S_m}(0) - E_{S_m}(\lambda_M)$. We here use the standard distinction between work and heat where work is considered as the change in eigenvalues of a quantum system, while heat is the redistribution of occupation probabilities among those eigenvalues,~\cite{alicki1979quantum, kieu2004second} and we define energy flow out from the double QDs to be positive.

The step from B to C is achieved by waiting long enough that the system thermalizes, meaning that its state becomes a classical thermal mixed state of $S_m$ and (the three-fold degenerate) $T$ described by the probabilities $P(T,\lambda_M) = 3 / (3 + \mathrm{exp}[\delta E / k_B\tau])$ and $P(S,\lambda_M) = 1-P(T,\lambda_M)$, where $\delta E = E_T(\lambda_M) - E_{S_m}(\lambda_M)$. This thermalization corresponds to absorbing the heat $Q_{BC} = - \delta E \times P(T,\lambda_M)$ from the bath.

\begin{figure}[hbt!]
	\centering
	\includegraphics[width=0.42\textwidth]{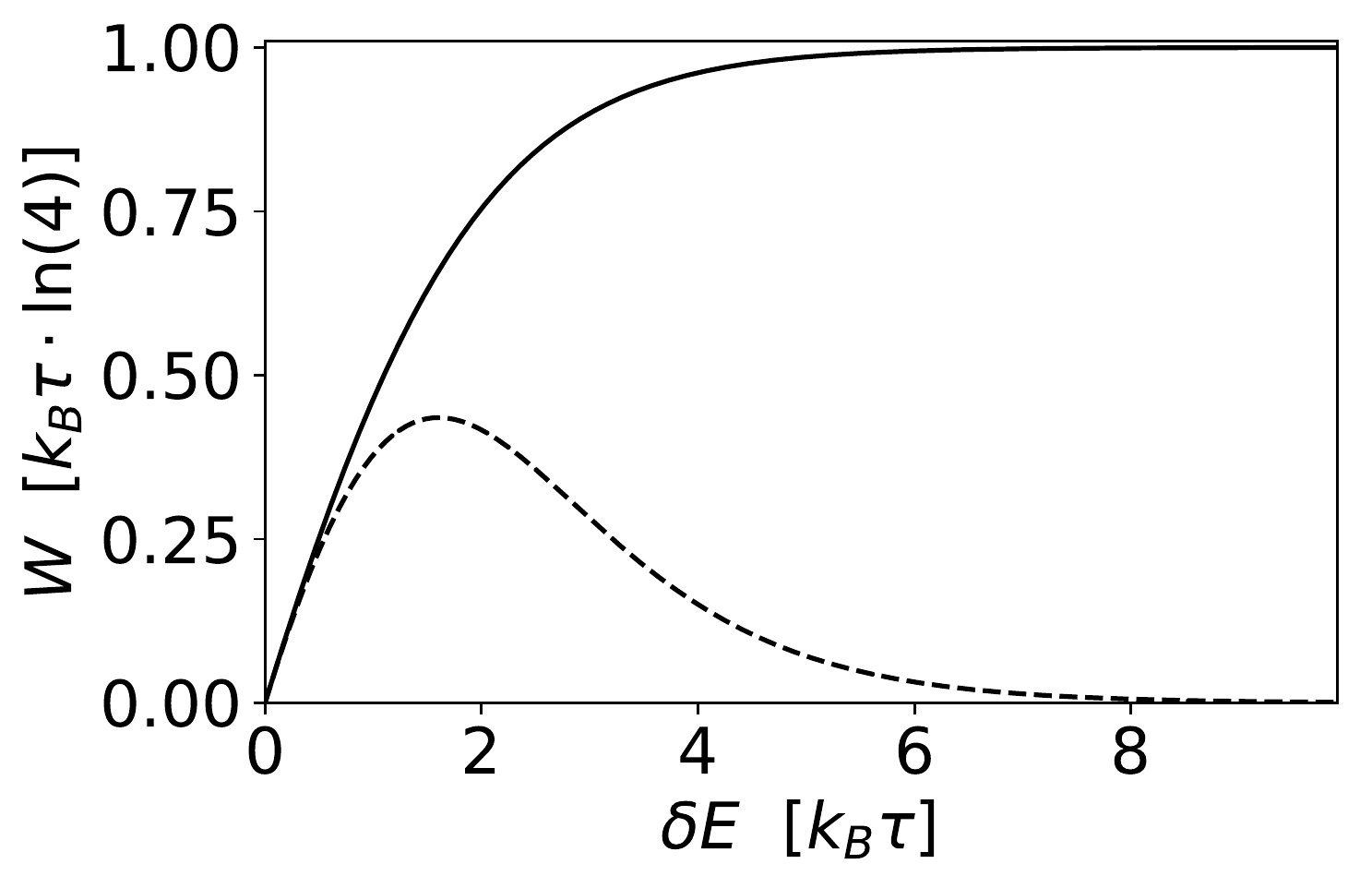}
	\caption{Work output as a function of the singlet-triplet energy split for cycle {\it (i)} (dashed line) and cycle {\it (ii)} (solid line).}		
	\label{fig:ideal}
\end{figure}

Finally, we move back from C to A and we distinguish between two different ways of doing this step. {\it (i)} If this final step is done adiabatically, we will on average have to supply the work $W^{(i)}_{CA} = -P(S,\lambda_M) [E_{S_m}(0) - E_{S_m}(\lambda_M)]$. Clearly, $W^{(i)} = W_{AB}+W_{CA}^{(i)} > 0$ and we have extracted work from the system. Furthermore, the first law gives that $W^{(i)} = -Q_{BC}^{(i)}$. {\it (ii)} If the final step is instead done isothermally the electrons will always be in equilibrium with the environment. The work required to bring the system from C to A is then calculated using $W_{CA}^{(ii)}=-\sum_n \int_{\lambda_M}^0 P_n\frac{d E_n}{d \lambda}d\lambda$, where the probabilities $P_n$ are given by a thermal state for all $\lambda$. This yields $W_{CA}^{(ii)} = k_B\tau\left[ \ln 4 -\ln Z(\lambda_M)\right],$ where $Z(\lambda_M) = 3+\exp[\delta E/k_B\tau]$, and a total amount of work $W^{(ii)} = W_{AB}+W_{CA}^{(ii)} = -(Q_{BC} + Q_{CA}^{(ii)} )> 0$ is extracted per cycle. In both cycles {\it (i)} and {\it (ii)} we thus extracted heat from a single bath and converted it perfectly into (electrical) work using the entropy (information) difference between the initial and final states. For the cycle to be repeated, a new input $S_m$ state has to be supplied or created.  
\\

The net work output, $W$, depends on $\delta E$ as can be seen in Fig.~\ref{fig:ideal}. In the adiabatic cycle {\it (i)} $W^{(i)}$ initially increases with the energy split, but once $\delta E \gtrsim k_B\tau$ the probability to end in $T$ becomes increasingly small and $W^{(i)}$ vanishes for big $\delta E$. In contrast, the work output in the isothermal cycle {\it (ii)} is maximal for large $\delta E$, for which $Q_{BC}\rightarrow 0$ and 
\begin{equation}
	W=\lim_{\delta E\rightarrow \infty}W_{AB}+W_{CA}=\delta E + k_BT\ln 4 - \delta E=k_B\tau\ln 4. 
\end{equation}
In this case the system converts an amount of heat equal to the free energy difference of the initial and final states, $\Delta F=\Delta E - T\Delta S$, into work, and thus acts as an ideal four-level engine. The limit of large $\delta E$ is equivalent to merging points $B$ and $C$ in Fig.~\ref{fig:setup}(b) (since $Q_{BC}=0$) and ideal performance is found when $A \rightarrow BC$ is performed adiabatically while $BC\rightarrow A$ is done in a quasi-static manner. However, even though the work output from cycle {\it (i)} is always lower than cycle {\it (ii)} under ideal conditions, the adiabatic cycle can still be preferred since it can be performed faster, possibly yielding a larger power output as we will show next. 
\section{\label{sec:details} Performance under realistic experimental conditions}
In this section we investigate the performance of the two cycles for a realistic (non ideal) engine. Concretely, we include a microscopic description of the thermal bath and a finite cycle time leading to the state after interaction with the bath not being a perfect thermal state. We assume that spin-orbit coupling and/or magnetic interactions with the environment cause spin flips of the individual electrons on the double QD where the energy difference can be absorbed by or emitted to a phonon bath,
\begin{equation}
	H_{B} = \sum_\omega  \varepsilon_k b^\dagger_k b_k + \sum_{i,k}\gamma c^\dagger_{i,\sigma}c_{i,\bar{\sigma}}(b^\dagger_k + b_k) + h.c.,
\end{equation}
where $b^\dagger_k\ (b_k)$ creates (annihilates) a phonon in the bath. The bath is assumed to be in a thermal state described by the Bose-Einstein distribution $n(\varepsilon_k)=(\exp [\varepsilon_k/k_B\tau] -1)^{-1}$ at all times. We take the phonon density of states to be linear $\nu(\varepsilon_k)=\nu_0 \varepsilon_k$, which is expected to hold for small energies,~\cite{danon2013spin} but note that 
our results would remain qualitatively the same also for other forms of $\nu(\varepsilon_k)$.
\\

In order to make predictions about relevant time scales and power output we include an explicit time dependence of $\lambda$, and the total work extracted from the system during a cycle is~\cite{alicki1979quantum}
\begin{equation}
	W = -\int_{t_{0}}^{t_{final}}\text{Tr}\left [\frac{dH_{DQD}(t)}{dt}\rho(t)\right ]dt,
\end{equation}
where $\rho(t)$ denotes the time-dependent density matrix of the double QD. We obtain $\rho(t)$ by assuming a weak system-bath coupling $(\gamma)$ allowing us to describe the system using master equations on Lindblad form $(\hbar=1)$
\begin{equation}
	\begin{split}
		\dot{\rho} = -i[H_{DQD}(t),\ & \rho(t)] + \sum_{i,\sigma} \Big [ A_{i,\sigma}(t)\rho(t)A^\dagger_{i,\sigma}(t) - \\
		&\frac{1}{2}\{A_{i,\sigma}^\dagger(t) A_{i,\sigma}(t),\rho(t)\} \Big ].
	\end{split}
	\label{eq:Lindblad}
\end{equation}
Equation~(\ref{eq:Lindblad}) is solved using the time-dependent Lindblad equation solver included in the \verb|Python| package QuTiP.~\cite{johansson2012qutip, johansson2013qutip} The jump operators $A_{i,\sigma}$ for the different spin flip processes, expressed in the time-dependent eigenbasis $|a_t\rangle$  of $H_{DQD}(t)$, are given by $A_{i, \sigma} = \sum_{a,a'}|a'_t\rangle\sqrt{\gamma\nu(\left |E_{a'a}(t)\right|)\cdot (n(|E_{a'a}(t)|) + \theta(E_{a'a}(t)))}\times {\langle a'_t|}c^\dagger_{i,\sigma} c_{i, \bar{\sigma}}|a_t\rangle \langle a_t|$,~\cite{kirvsanskas2018phenomenological} where $E_{a'a}=E_{a'}-E_a$ and $\theta$ denotes the Heaviside step-function. In order to solve Eq.~(\ref{eq:Lindblad}) we further assume that all timescales are much larger than the bath correlation time such that memory effects in the bath can be neglected. We start and end the cycle at a small $\lambda_0>0$, which should be large enough that the minimum energy split between $S_m$ and $T$ is larger than the maximal matrix element of $A_{i,\sigma}(\lambda_0)$ connecting $S_m$  and $T$ in order for the secular approximation to be valid.~\cite{bulnes2016quantum} However, for weak system-bath couplings $\lambda_0$ can become arbitrary small and the effect of not setting $\lambda_0=0$ will be minuscule.
\\

We define the time dependence of the two cycles as follows. Both cycles start with increasing $\lambda(t)$ from $\lambda_0$ to $\lambda_M$ at constant rate during time $t_1$. In $(i)$ the system is kept at $\lambda_M^{(i)}$ during time $t_2^{(i)}$ until it is brought back to $\lambda_0$ at a constant rate during time $t_3=t_1$. In $(ii)$ the system is brought back to $\lambda_0$ at a constant rate during time $t_2^{(ii)}$. Typically $t_2\gg t_1$ and the ideal cases discussed in the previous section correspond to $t_1=0$ and $t_2=\infty$.
\begin{figure}[hbt!]
	\centering
	\includegraphics[width=0.48\textwidth, trim={7 0 0 0},clip]{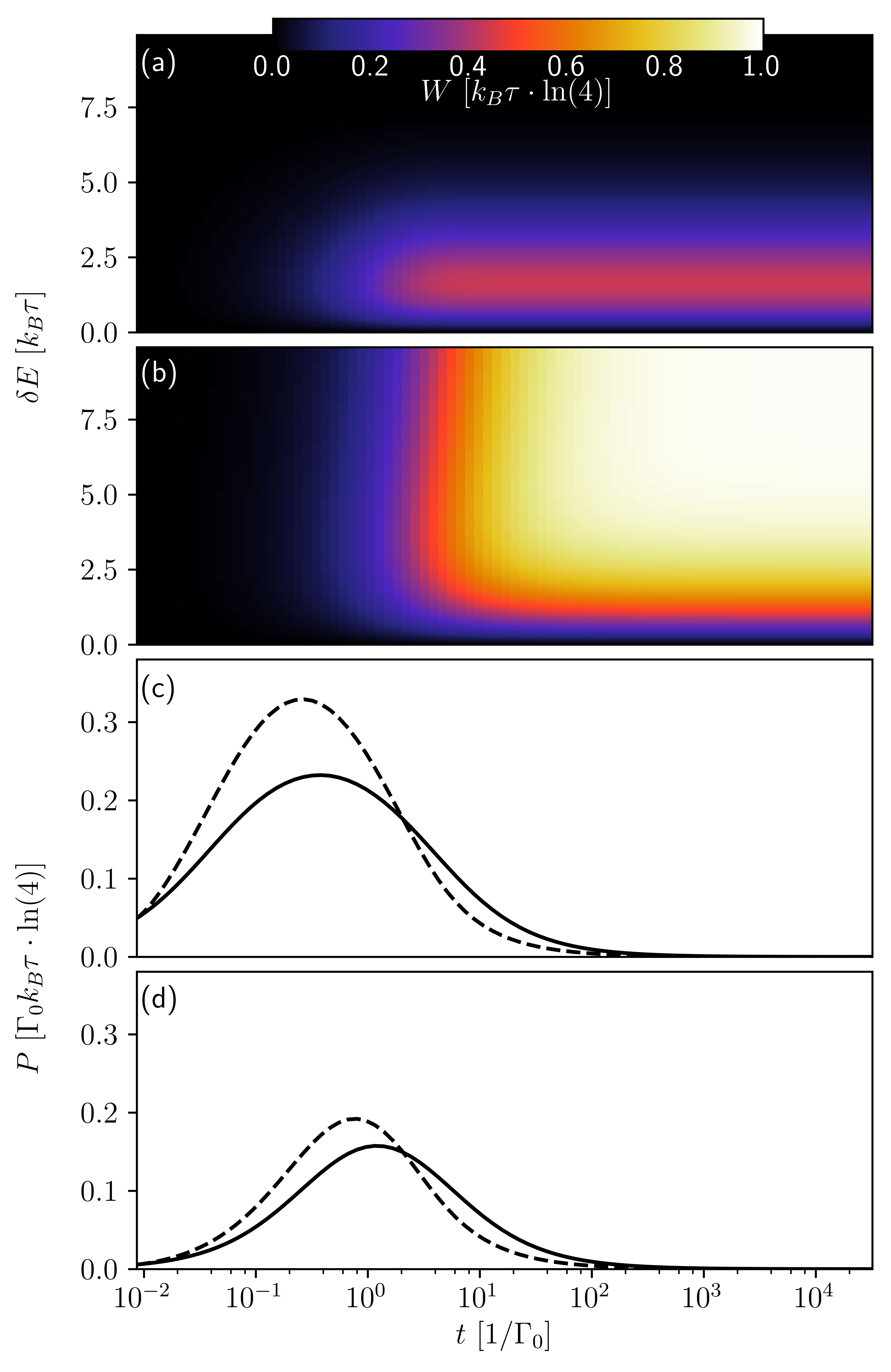}
	\caption{Work output per cycle for cycle {\it (i)} (a) and cycle {\it (ii)} (b) plotted against the maximum energy split $\delta E$ and total cycle time $t$. (c)-(d) show power output optimized w.r.t. $\delta E^{(i)/(ii)}$ as a function of the total cycle time excluding the singlet initialization time $t_{init}$ for cycle {\it (i)} (dashed line), and cycle {\it (ii)} (solid line). $\gamma$ is chosen such that the coupling between different $T$ states is $\hbar\Gamma_0/k_B\tau=\hbar\gamma \nu_0 = 3.2\cdot 10^{-4}$, and $E_T(\lambda_0)-E_S(\lambda_0)=1.6\cdot 10^{-3}k_B\tau$. $t_1$ is set to $0.004/\Gamma_0$, which corresponds to $1$ ns if $\tau=0.1$ K. Other parameters used are $U_{L}=U_R=100k_B\tau$ and $U_{LR} = 0$. The initial state is $S(1,1)= \frac{1}{\sqrt{2}}(c^\dagger_{1\uparrow}c^\dagger_{2\downarrow} - c^\dagger_{1\downarrow} c^\dagger_{2\uparrow})|0\rangle$. Additionally, in the power plots: (c) $t_{init}=0.04/\Gamma_0$ and (d) $t_{init}=0.4/\Gamma_0$. }
	\label{fig:WorkPower}
\end{figure}
\\

Figures~\ref{fig:WorkPower}(a)-(b) show how $W$ depends on $\lambda_M$ through $\delta E$ and the total cycle time using experimentally relevant parameters (see figure caption for details). A common feature for both cycles is the vanishingly small work output for short times as the system does not have enough time to transition to a triplet state. In order to be able to extract a significant amount of work the cycle time should be at least on the same order as the transition time between $S_m$ and $T$ induced by the bath, and for long times it should be possible to come arbitrarily close to $k_B\tau \ln 4$ using cycle $(ii)$.~\cite{browne2014guaranteed}
However, in a real life operating scenario, e.g. recycling of singlets from quantum computational operations, fast cycle times and larger power output can trump a large amount of work per cycle. We explore the power the engine can deliver to the gates for the two cases. To effectively represent more realistic performance we  also include a time $t_{init}$ for initialization of the qubit. The total cycle time then becomes $t_{init}+2t_1+t_2^{(i)}$ for cycle $\it(i)$ and $t_{init}+t_1+t_2^{(ii)}$ for cycle $\it(ii)$. The output power in Figs.~\ref{fig:WorkPower}(c)-(d) (maximized w.r.t. $\delta E^{(i)/(ii)}$) turns out to be larger in cycle {\it (i)} for almost all relevant times. The maximum power of $P\approx 0.3 \Gamma_0 k_B\tau\ln 4$ in Fig.~\ref{fig:WorkPower} corresponds to $2.5$ aW if $\tau=0.1$ K, $\Gamma_0=4.2$ MHz, $t_1=1$ ns. The reason cycle {\it (ii)} requires more time than cycle {\it (i)} to generate the same amount of work is that the system should re-thermalize for all infinitesimal changes in $\lambda$ to provide optimal work output in cycle {\it (ii)}, whereas the final step should be performed as fast as possible to hinder further interactions with the bath in cycle {\it (i)}. The power naturally depends on the time it takes for a new singlet to be provided, something that will vary depending on implementation and  experimental setup. But as a rule of thumb, cycle {\it (i)} provides more work per cycle for $t<\frac{1}{\Gamma_0}$ and cycle {\it (ii)} is preferable when $t>\frac{1}{\Gamma_0}$. As a result, picking the best cycle becomes a matter of characterizing the transition time between (almost) degenerate states. 

We note that other protocols for changing $\lambda(t)$ might provide better performance,\cite{suri2018speeding, cavina2018optimal, scandi2019thermodynamic, menczel2019two} leaving room for further optimizations. But the two cases presented here are the most straight-forward to implement in an experiment, and thus natural places to start. 

\section{\label{sec:conclusion} Conclusions}
We have proposed a thermodynamic quantum engine based on a double QD that extracts energy from a single thermal bath. We envision the engine being powered by a constant flow of two-electron singlet states, a condition that may be met in future quantum computers or simulators. The performance is studied for two cycles under both ideal and realistic experimental conditions. In an ideal quasi-static case, complete knowledge of the pure, maximally entangled, initial state allows for converting a maximum of $k_B\tau\ln 4$ of thermal energy to work per cycle at the expense of increasing the entropy of the electrons on the double QD. The final mixed state after a cycle has exactly the same energy as the initial maximally entangled state and any measurement on a single electron spin would be unable to differentiate between the two. Under realistic, finite time, conditions we find an alternative cycle to be superior in terms of power output whenever the cycle time is shorter than the environment-induced transition at small energy gaps.
\\

The engine is compatible with the well established platform of semiconductor QD qubits, making an experimental realization in the near future possible. Ideally, an experimenter would measure the generated power, which, unfortunately, is a highly non-trivial task. Instead, a proof-of-principle experiment could be conducted by performing measurements of the quantum state at different times during the cycle to map out the time-dependent occupations of the singlet and triplet states,\cite{petta2005coherent, barthel2009rapid} or by measuring the heat flow out from a mesoscopic reservoir. 
\section*{Acknowledgements}
We thank Patrick Potts, Peter Samuelsson and Ville Maisi for providing valuable feedback on the manuscript. We acknowledge funding from NanoLund and the Knut and Alice Wallenberg Foundation (project 2016.0089). Computational resources were provided by the Swedish National Infrastructure for Computing (SNIC) at LUNARC.
\section*{APPENDIX}
\appendix
The eigenstates for two electrons on the DQD system at $\lambda=0$ are
\begin{equation}
	\begin{split}
	&|S_0\rangle = S(1,1)= \frac{1}{\sqrt{2}}(c^\dagger_{1\uparrow}c^\dagger_{2\downarrow} - c^\dagger_{1\downarrow} c^\dagger_{2\uparrow})|0\rangle, \\ 
	&|T_0\rangle = \frac{1}{\sqrt{2}}(c^\dagger_{1\uparrow}c^\dagger_{2\downarrow} + c^\dagger_{1\downarrow} c^\dagger_{2\uparrow})|0\rangle, \\
	&|S_+\rangle = \frac{1}{\sqrt{2}}(c^\dagger_{1\uparrow}c^\dagger_{1\downarrow} + c^\dagger_{2\uparrow}c^\dagger_{2\downarrow})|0\rangle, \
	|T_+\rangle = c^\dagger_{1\uparrow}c^\dagger_{2\uparrow}|0\rangle, \\
	&|S_-\rangle = \frac{1}{\sqrt{2}}(c^\dagger_{1\uparrow}c^\dagger_{1\downarrow} - c^\dagger_{2\uparrow}c^\dagger_{2\downarrow})|0\rangle, \
	|T_-\rangle = c^\dagger_{1\downarrow}c^\dagger_{2\downarrow}|0\rangle.
\end{split}
\end{equation}
When analyzing the engine performance we set $\epsilon_L=\epsilon_R=U_{LR}=0$ and restrict our analysis to the subspace spanned by the four lowest states, which for $\lambda>0$ are all $T$s as well as 
\begin{equation}
	S_m = \frac{1}{\sqrt{E_{S_m}^2+4\lambda^2}}\big ( 2\lambda|S_0\rangle + E_{S_m}|S_+\rangle\big ).
\end{equation}
The two remaining states are separated from the others by an energy split $\Delta E\approx U=U_{L}=U_{R}\gg k_B\tau$ (we take the intra-QD interactions to be equal), and transitions to these states are very unlikely. The energies in the relevant subspace are
\begin{align}
	E_T(\lambda)=0,\ E_{S_m}(\lambda)=\frac{U}{2} - \sqrt{\frac{U^2}{4}+4\lambda^2}.
\end{align} 
\vspace{7cm}
\bibliographystyle{apsrev_modified}
\bibliography{refs.bib}
\end{document}